# Robust LSB Watermarking Optimized for Local Structural Similarity

Amin Banitalebi, Said Nader-Esfahani, Alireza Nasiri Avanaki
School of Electrical and Computer Engineering
University of Tehran, Tehran 14395-515, Iran
banitalebi@ieee.org

*Abstract*—Growth of the Internet and networked multimedia systems has emphasized the need for copyright protection of the media. Media can be images, audio clips, videos and etc. Digital watermarking is today extensively used for many applications such as authentication of ownership or identification of illegal copies. Digital watermark is an invisible or maybe visible structure added to the original media (known as asset). Images are considered as communication channel when they are subject to a watermark embedding procedure so in the case of embedding a digital watermark in an image, the capacity of the channel should be considered. There is a trade-off between imperceptibility, robustness and capacity for embedding a watermark in an asset. In the case of image watermarks, it is reasonable that the watermarking algorithm should depend on the content and structure of the image. Conventionally, mean squared error (MSE) has been used as a common distortion measure to assess the quality of the images. Newly developed quality metrics proposed some distortion measures that are based on human visual system (HVS). These metrics show that MSE is not based on HVS and it has a lack of accuracy when dealing with perceptually important signals such as images and videos. SSIM or structural similarity is a state of the art HVS based image quality criterion that has recently been of much interest. In this paper we propose a robust least significant bit (LSB) watermarking scheme which is optimized for structural similarity. The watermark is embedded into a host image through an adaptive algorithm. Various attacks examined on the embedding approach and simulation results revealed the fact that the watermarked sequence can be extracted with an acceptable accuracy after all attacks. In comparison to the original algorithm, the proposed method increases the capacity while the imperceptibility and robustness remain fix.

*Keywords-digital watermarking; structural similarity; Least significant bit watermarking.*

## I. INTRODUCTION

Watermarking is usually defined as the practice of imperceptible altering a work to embed a message about that work [6]. With the growth of the Internet and multimedia industry the usage of the watermarking for proof of ownership and copyright protection is an indispensable necessity.

There are some considerations when embedding a digital watermark into a host signal. The main concern of the embedding part of any data hiding system is to make the hidden data imperceptible. This task can be achieved either implicitly, by properly choosing the set of host features and embedding rule, or explicitly, by introducing a concealment step after embedding [7]. But, there is trade-off between imperceptibility, robustness and capacity of the watermarking system; a triangular trade-off in which moving toward each corner will take us away from the other corners. Perhaps we can introduce another important parameter, the complexity of the embedding algorithm or speed of each embedding action. This parameter is important when embedding action should be done rapidly and maybe iteratively for each user. Transaction watermarking is the best example for this idea [5]. Transaction watermarking is copyright protection for online shops by customer identification watermarking in which a specific watermark should be embedded in the media each time a user downloads that media. An ideal watermarking scheme should considers these trade-offs into account in the best way.

In the case of image signals, watermark can directly be embedded to the spatial domain or in the transform domain. It can be visible or invisible. Here, we address the invisible watermarking strategies. As we mentioned above, there are two main requirements of invisible watermarks: Embedded watermark should be perceptually invisible and robust to common signal processing and intentional attacks while the capacity is already taken into account. Focus of early researches on digital watermarking was almost on the first objective without considering the second one. Recently much works have been done to find the best trade-off between robustness, imperceptibility and capacity [8-14]. In some methods watermarks are implied in the pixel domain [15-17]. Embedding in the asset domain could have possible advantages such as low complexity, low cost, low delay as well as this fact that temporal/spatial localization of the watermark is automatically achieved, thus permitting a better characterization of the distortion introduced by the watermark and its possible annoying effects. In the transform domain, the watermark is inserted into the coefficients of a digital transform of the host asset. Barni's main work was on the discrete cosine transform (DCT) and discrete Fourier transform (DFT) [20]. Embedding in the wavelet domain transform is also of much interest [2]. [18] Utilized DCT and [19] utilized contourlet as their transforms. Usually transform domain watermarks exhibit a higher robustness to the attacks. More imperceptibility can be achieved by avoiding the changes into the low frequency components. Complexity of this type of embedding can be the trade-off of this system.

It has been showed that MSE cannot track the human perception carefully [21-23]. The Applications of HVS based image quality criterions are being spread recent years. Some works have been done in the field of digital watermarking with the human perception concluded in those approaches [2,3,4]. For example, [3] used a noise visibility function (NVF) to set a threshold for embedding rule capacity.

In this paper we propose a novel digital watermarking method that utilizes SSIM in the embedding and extraction rules. We have tested our algorithm on the LSB watermarking scheme. The original LSB watermarking is modified so that the capacity is maximized, while the watermark remains imperceptible. Insertion of the watermark sequence is an adaptive procedure. The stop condition for watermark insertion in each block of image is specified by an SSIM threshold.

In section II SSIM theory will be review. Section III describes our proposed method and section IV is dedicated to the simulation results as well as comparison while section V concludes the paper.

## II. STRUCTURAL SIMILARITY BASICS

Natural images are highly structured and this local dependency of the pixels is not completely removed during a digital transform. This fact is the basic idea to the definition of the SSIM. The reference (original) and the distorted images (**x**, **y** respectively) are decomposed to their luminance, contrast and structure components. These components are defined as:

$$l(x,y) = \left(\frac{2\mu_x \mu_y + C_1}{\mu_x^2 + \mu_y^2 + C_1}\right), \quad c(x,y) = \left(\frac{2\sigma_x \sigma_y + C_2}{\sigma_x^2 + \sigma_y^2 + C_2}\right)$$

$$, s(x,y) = \left(\frac{\sigma_{xy} + C_3}{\sigma_x \sigma_y + C_3}\right) \qquad (1)$$

And the local SSIM metric is defined as:

$$S(x,y) = l(x,y)^\alpha \cdot c(x,y)^\beta \cdot s(x,y)^\gamma = $$
$$\left(\frac{2\mu_x \mu_y + C_1}{\mu_x^2 + \mu_y^2 + C_1}\right)^\alpha \cdot \left(\frac{2\sigma_x \sigma_y + C_2}{\sigma_x^2 + \sigma_y^2 + C_2}\right)^\beta \cdot \left(\frac{\sigma_{xy} + C_3}{\sigma_x \sigma_y + C_3}\right)^\gamma \qquad (2)$$

In the above equation, $\alpha, \beta, \gamma$ are three parameters that are used for the adjustment of the importance of each of the three components. We suppose that they have all equal importance. $\mu_x$ and $\mu_y$ are (respectively) the local sample means of *x* and *y*, $\sigma_x$ and $\sigma_y$ are (respectively) the local sample standard deviations of **x** and **y**, and $\sigma_{xy}$ is the sample cross correlation of **x** and **y** after removing their means. The items $C_1$, $C_2$, and $C_3$ are small positive constants that stabilize each term, so that near-zero sample means, variances, or correlations do not lead to numerical instability. The entire SSIM metric between the original and the reference image is calculated by averaging the local SSIM all over the image. More details are available at [23].

## III. PROPOSED METHOD

### A. Original LSB algorithm

We will have a short review on original LSB approach. The asset is image **A** and the watermark image is **B**. N is a parameter that shows the embedding depth. Suppose that we have 8-bit images. The algorithm says that: N left-sided bits of the image **B** should be replaced with the N right-sided bits of the image **A** for each pixel (represented in 8-bit format). In this way, most significant bits or equally important information of image **B** is watermarked in the place of least significant bits or details of the image **A**. Fig. 1 shows the original Lena image (**A**) and watermark image (**B**) and corresponding watermarked image.

### B. Proposed Algorithm

The proposed algorithm finds a good trade-off point between capacity and imperceptibility and robustness. The embedding process can be done in the following steps:

- Step1: divide the asset (**A**) and the watermark (**B**) images into $k \times k$ blocks. We set *k=11* for the sake of simulation speed and complexity reduction.

- Step 2: for the *i*th block of both images, compute the SSIM index between these two blocks. Name this local block similarity as *SSIM_i*.

- Step 3: if *SSIM_i* is greater than a pre-specified threshold (*thr1*) then do the original LSB embedding algorithm for one bit of only these two blocks, i. e. replace the left sided bit of all pixels of the *i*th block of **B** with the right sided bit of corresponding pixels of the *i*th block of image **A**.

- Step 4: compute SSIM between the *i*th block of **A** and the *i*th block (the block generated in step 3) of the new image (watermarked image or **W**). If the similarity is greater than a second threshold (*thr2*, we chose 0.75) then repeat step 3 for another bit, i. e. embed another bit of **B** in **W**. For example if we have just embedded the *j*th bit of **B** in **W** and the new SSIM is bigger than *thr2* then we should embed the *(j+1)*th bit.

- Step 5: repeat step 4 until the new computed SSIM would become less than *thr2*. Stop the embedding in this block now and go to step 2 for the *(i+1)*th block.

The proposed algorithm leads us to a local optimum of watermark embedding in the case that by adjusting the threshold value one can directly tune the imperceptibility and capacity. Another advantage of this method is that it is robust to most of the attacks. As we will see in the results section, unlike the crude LSB algorithm, the proposed method can extract the true watermark sequence with an acceptable accuracy after various attacks or processing. The extraction procedure would be easy. If we could have the place and location of the embedded bits as a watermark key then we can easily extract the embedded sequence.

## IV. RESULTS AND COMPARISON

The proposed algorithm was simulated in MATLAB and Lena is chosen for asset and Cameraman as the watermark image. Fig. 2 shows the simulation results for *thr2=0.75* and different *thr1* values. As it can be seen from the fig. 2, *thr1=0.8* is a good point of the trade-off between imperceptibility and robustness and capacity, but, due to the application one can choose arbitrary thresholds. Table 1 sketches the response of the extraction of the watermark after processing and attacks. Error percentage is the Hamming distance between the extracted watermark and true watermark. As it is seen from the table, the proposed method is robust against the attacks while the original LSB approach is not. Fig. 4 demonstrates the detectors output to the 100 randomly generated sequences. As we can see, the detector shows the biggest correlation in the case of the true watermark sequence. In comparison to the original LSB algorithm, it is noteworthy that the proposed algorithm would require more computational complexity but we can achieve a better trade-off between capacity and imperceptibility and also a better robustness against the attacks and unavoidable artifacts will be available. The structural similarity based method offers an adaptive embedding procedure that will embed the watermark sequence until a threshold for imperceptibility is achieved. It is clear that less threshold values will results in a bigger capacity but less invisibility. Choosing *thr1* equal to zero makes the new algorithm just very similar to the original LSB method.

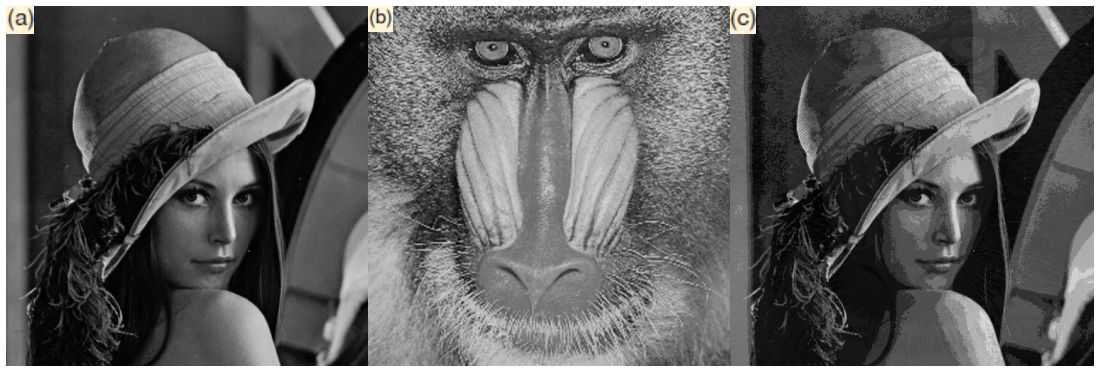

Figure 1: Original LSB algorithm; (a) Asset (b) Watermark Image (c) Watermarked Image

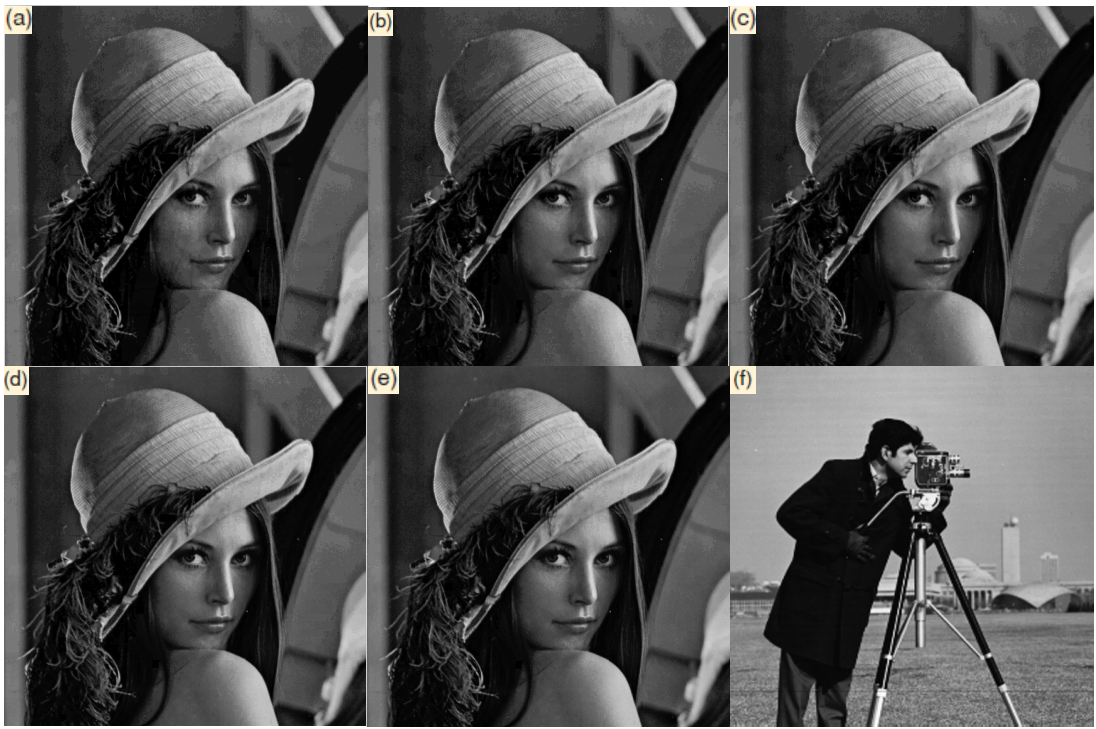

Figure 2: Watermarked Image resulted from the proposed algorithm; (a) Thr1=0 (b) Thr1=0.2 (c) Thr1=0.4 (d) Thr1=0.6 (e) Thr1=0.8 (f) Watermark Image

Table 1: Accuracy of the watermark extraction during various attacks: extraction error percentage

|  | Threshold1=0 | Threshold1=0.2 | Threshold1=0.4 | Threshold1=0.6 | Threshold1=0.8 | Original LSB |
|---|---|---|---|---|---|---|
| Motion Blur | 0.196 | 0.178 | 0.152 | 0.142 | 0.105 | 0.288 |
| Jpeg Compression | 0.143 | 0.124 | 0.108 | 0.089 | 0.054 | 0.225 |
| Low Pass Filtering | 0.154 | 0.142 | 0.122 | 0.114 | 0.073 | 0.239 |
| Cropping | 0.025 | 0.020 | 0.015 | 0.011 | 0.002 | 0.025 |
| Gaussian Noise | 0.230 | 0.213 | 0.178 | 0.152 | 0.116 | 0.268 |
| Salt & Pepper Noise | 0.213 | 0.184 | 0.155 | 0.123 | 0.094 | 0.296 |

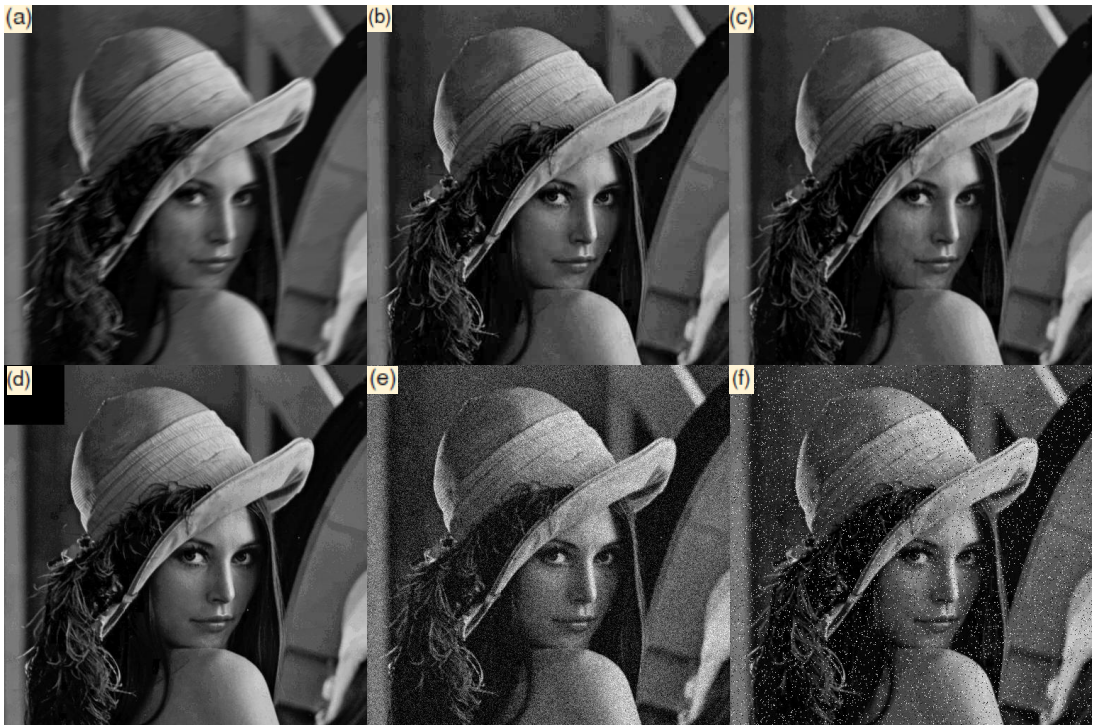

Figure 3: Watermarked image of proposed method after various attack types: (a) motion blur (b) jpeg compression (c) low pass filtering (d) cropping (e) Gaussian noise (f) salt & pepper noise

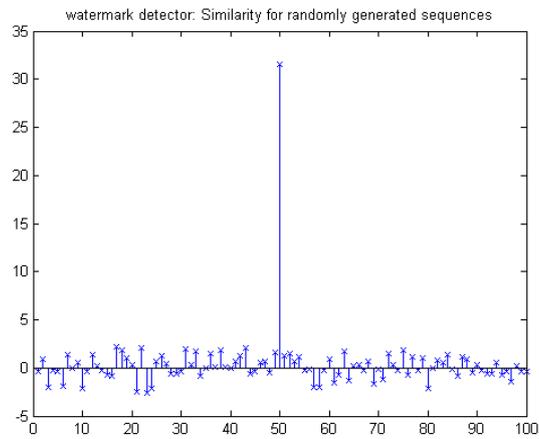

Figure 4: Watermark detector for 100 randomly generated sequences

## V. CONCLUSION

In this paper we first gave a historical review on digital watermarking and its trade-offs and then we state the necessity of the usage of HVS based quality metrics. A short survey on SSIM was reviewed. We proposed a new modified LSB watermarking technique based on structural similarity. Simulation results proved that the proposed method can find a better trade-off between capacity and imperceptibility than the original version. Robustness of the new algorithm was examined during the attacks. Results revealed this fact that the watermarked sequence can be extracted with an acceptable accuracy after all attacks.